\documentclass[%
 reprint,
superscriptaddress,
 amsmath,amssymb,
 aps,
]{revtex4-1}

\usepackage{graphicx}
\usepackage{dcolumn}
\usepackage{bm}

\usepackage{color}
\usepackage{braket}
\usepackage{dsfont}
\usepackage{isotope}
\usepackage{multirow}
\usepackage{hyperref}

\usepackage{tikz-feynman}

\newcommand{\hQ}{\hat{Q}}
\newcommand{\hR}{\hat{R}}
\newcommand{\hP}{\hat{P}}
\newcommand{\hC}{\hat{C}}
\newcommand{\hG}{\hat{G}}
\newcommand{\hH}{\hat{H}}
\newcommand{\hV}{\hat{V}}
\newcommand{\hPhi}{\hat{\Phi}}
\newcommand{\hphi}{\hat{\phi}}
\newcommand{\hLambda}{\hat{\Lambda}}
\newcommand{\hT}{\hat{T}}
\newcommand{\hU}{\hat{U}}
\newcommand{\hW}{\hat{W}}
\newcommand{\hY}{\hat{Y}}

\begin{document}

\title{Theory of Laser-Assisted Nuclear Excitation by Electron Capture}

\author{Pavlo Bilous}
\email{pavlo.bilous@mpl.mpg.de}
\affiliation{Max Planck Institute for the Science of Light, Staudtstr. 2, 91058 Erlangen, Germany}

\date{\today}

\begin{abstract}
The interplay of x-ray ionization and atomic and nuclear degrees of freedom is investigated theoretically in the process of laser-assisted nuclear excitation by electron capture.  In the resonant process of nuclear excitation by electron capture,  an incident electron recombines into 
 a vacancy in the atomic shell with simultaneous nuclear excitation. Here we investigate the specific scenario in which the free electron and the required atomic shell hole are generated by an x-ray free electron laser pulse.  We develop a theoretical description based on the Feshbach projection operator formalism and consider  numerically experimental scenarios at the SACLA x-ray free electron laser.  Our numerical results for excitation of the 29.2~keV nuclear state in $\isotope[229]{Th}$ and the 14.4 keV M\"ossbauer transition in $\isotope[57]{Fe}$  show low excitation rates but strong enhancement with respect to direct two  photon nuclear excitation.
  
\end{abstract}

\maketitle

\section{Introduction}

Typical energies of electromagnetic transitions in nuclei span over a wide range from a few keV to hundreds MeV.  Transitions at the lower energy limit are of special interest due to their narrow width of $10^{-7}$ down to $10^{-11}$~eV or less.  The main application of this property, in conjunction with recoilless photon absorption and reemission, has been the method of M\"ossbauer spectroscopy widely used in material science, geology, chemistry and biology \cite{Moessbauer1958,  Cadogan_Mossbauer_2006,  Yoshida_ModernMoessbauer2021}.
Nuclear excitation can occur via photon absorption,  using either radioactive M\"ossbauer sources as in the original experiment of M\"ossbauer \cite{Moessbauer1958},  or synchrotron radiation in nuclear forward scattering or grazing incidence \cite{Smirnov_MoessbauerReview_1986},  or most recently radiation from x-ray free electron lasers (XFEL) \cite{Chumakov2018}. The commissioning and operation of the first XFELs with photon energies up to 20 keV at the SACLA facility \cite{SACLA_overview} benefit the  emerging field of x-ray  quantum optics \cite{Adams2013}. 

An alternative and less investigated possibility to address low-lying nuclear transitions is the mechanism of nuclear excitation by electron capture (NEEC) theoretically considered e.g.  in Refs. \cite{Goldanskii,Palffy_NEECTheory_PRA_2006}. In the resonant process of NEEC, an incident electron recombines into 
 a vacancy in the atomic shell with simultaneous nuclear excitation. This is the time-reversed process of internal conversion (IC), in which nuclear excitation is not released with an irradiated photon but is transferred to an atomic electron, which leaves the atom. Just recently, NEEC has been experimentally observed \cite{NEEC_Nature_2018}, giving rise to quite some controversy in following theoretical and experimental works \cite{Rzadkiewicz2021_PRL, Guo2022_PRL, Wu_NEEC_Mo_PRL_2019, Guo2021_Nature}. 
 
Typically, the free electrons required in the process  of NEEC are obtained for instance from laser-generated plasmas \cite{Gunst2014_PRL,  Gunst_PhysRevE.97.063205,  Wu2018_PRL, Wu_PRA_2019} or passing the atoms through a solid target \cite{NEEC_Nature_2018, Rzadkiewicz2021_PRL,Guo2022_PRL}.  
In this work we consider a different possibility in an extension of the NEEC process,  in which the impact electron stems from the atomic cloud surrounding the nucleus to be excited.  Expelling of this electron from the electronic shell is achieved by an x-ray photon generated at an XFEL facility.  The continuum electron is then captured to a vacant bound state with simultaneous excitation of the nucleus.  We refer to this process as to laser-assisted nuclear excitation by electron capture (LANEEC) proposed for the first time in Ref.~\cite{BilousThesis2018}.  In this simplest form involving one external photon,  LANEEC has been recently considered in Ref. \cite{Borisyuk2019_PRC} (denoted by the authors as ``electronic bridge excitation via continuum'') for excitation of the $\isotope[229m]{Th}$ nuclear isomer using an optical laser.  The process has been theoretically described by means of perturbation theory \cite{Borisyuk2019_PRC} and scattering theory \cite{Dzyublik2020_PRC}.  In this work we consider LANEEC with one or two x-ray photons based on the Feshbach projection operator method in the form described in Ref. \cite{HaanJacobs1989} which provides a unified description to all orders for the decay channels of the involved states. 

\begin{figure}[ht!]
\centering
\includegraphics[width=0.5\textwidth]{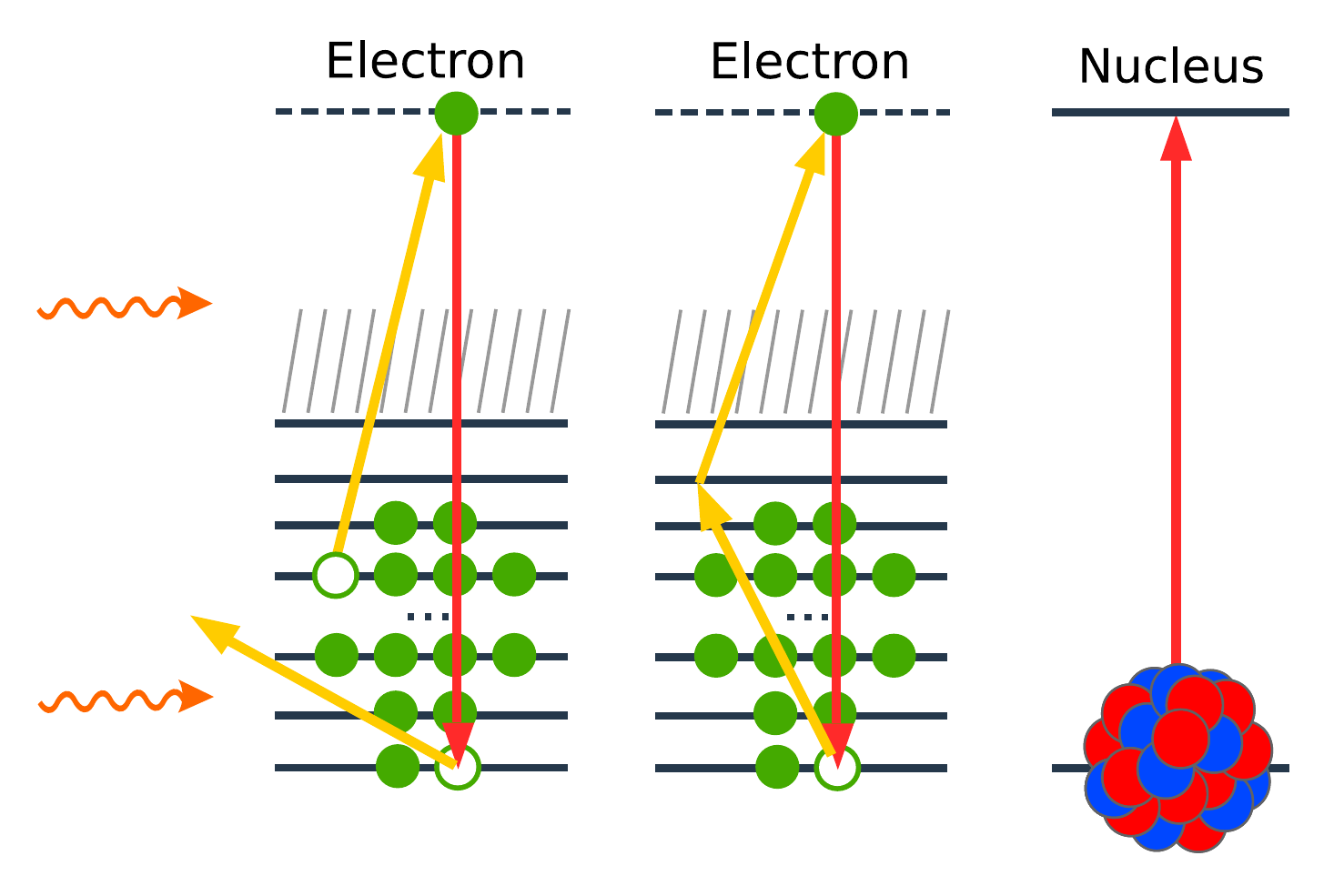}
\caption{
(Color online) Schematic illustration of two LANEEC versions with two photons.  The nuclear transition in both cases is shown in the right graph (red arrow).  In the first version referred to as ``LANEEC with additional hole'' (left graph),  an x-ray creates a vacancy in an inner shell by expelling an electron (lower yellow arrow).  Another photon induces the ordinary one-photon LANEEC with final electronic state in the created vacancy (upper yellow and red arrows).  In the second version referred to as ``two-photon LANEEC'', (middle graph) an inner-shell electron absorbs two photons and is promoted to a continuum state (yellow arrows) which decays to the created vacancy (red arrow) with excitation of the nucleus.
\label{2photLANEEC_versions}}
\end{figure}

In the simplest LANEEC scenario with one x-ray photon,  the electronic path starts from a fully occupied inner shell and ends in a vacant outer shell.  The electron partially takes over the energy carried by the exciting photon and the latter has to be therefore larger than the nuclear transition energy.  Apart from this ``pure'' LANEEC, we consider in this work two improved LANEEC versions with an additional x-ray photon.  They both allow usage of photons at lower energies and thus addressing nuclear transitions with energies lying beyond the range achievable today at XFEL facilities. The two considered scenarios are depicted in Fig.~\ref{2photLANEEC_versions}.  The nuclear transition in both cases is shown in the right graph (red arrow).  The electronic part in the first LANEEC version is depicted in the left graph.  Here an x-ray photon expels an electron from a deep-lying inner shell creating a vacancy (lower yellow arrow).  Another photon promotes another atomic electron into a continuum state and induces the ordinary one-photon LANEEC process with final electronic state in the vacancy created by the first photon (upper yellow and red arrows).  In the second scenario with the electronic part depicted in the middle graph,  an inner-shell electron absorbs two photons and is promoted to a continuum state (yellow arrows) with further capture to the created vacancy (red arrow).  Here the first excitation step and the one-photon LANEEC step are combined at the level of amplitudes to yield the amplitude of the compound LANEEC process.  In order to distinguish between the two processes, we refer to them in the following as ``LANEEC with additional hole'' and ``two-photon LANEEC'', respectrively, stressing the fact, that the electron involved directly in LANEEC experiences a two-photon transition only in the latter version.

We consider concrete numerical examples for each of the three LANEEC processes in an experimental implementation at the SACLA facility \cite{SACLA_overview},  including excitation of the 29.2~keV nuclear state in $\isotope[229]{Th}$ and the 14.4 keV M\"ossbauer transition in $\isotope[57]{Fe}$.  The calculations show that due to small excitation rates the considered schemes are very challenging or impractical today but may be of interest for future applications.  The latter conclusion is supported by numerical results for the LANEEC versions with two photons, which show the rates many orders of magnitude larger than the rate of direct excitation of the same nuclear transition with two photons. 

The article is structured as follows.  In Section~\ref{Sec_ME} we apply the Feshbach projection operator formalism and derive a general expression for the matrix element of the transition operator describing the LANEEC excitation with one incident photon.  In Section~\ref{Sec_1phot} we obtain expressions for the amplitude and the rate of the aforementioned process.  The obtained rate is adopted also for LANEEC with an additional hole.  In Section~\ref{Sec_2phot} we obtain the expressions for the amplitude and the rate of two-photon LANEEC. In Section~\ref{Sec_NR} we consider concrete numerical examples for each LANEEC scenario.  In the final Section~\ref{Sec_Concl} we discuss conclusions following from the obtained results.  Atomic units $\hbar = e = m_e = 1$ are used unless otherwise stated.

\section{Matrix element of transition operator\label{Sec_ME}}

In this section we develop a very general theoretical description of the LANEEC process based on the Feshbach projection operator formalism.  According to this approach, we separate the total Hilbert space of the system states into mutually orthogonal subspaces $Q$, $R$ and $P$. The subspace $Q$ consists of the states with all electrons bound in the atomic shell and no photons present; states from $R$ describe the system with all electrons bound plus one photon; $P$ consists of the states with no photons, one electron in a continuum state and the other electrons bound. We denote the basis states in $Q$, $R$ and $P$ as $\ket{\beta}$, $\ket{f \omega}$ and $\ket{\alpha \varepsilon}$, respectively, where the first symbol denotes the set of all discrete quantum numbers characterizing the state, and the second one (for the states from $R$ and $P$) is the total energy of the system, which is a continuous quantity due to presence of a photon or a continuum electron.  We assume the following normalization
\begin{eqnarray}
\braket{\beta | \beta'} &=& \delta_{ \beta \beta' } \;,\\
\braket{ f \omega | f' \omega' } &=& \delta_{f f'} \delta( \omega-\omega' )\;,\\
\braket{ \alpha \varepsilon | \alpha' \varepsilon'} &=& \delta_{\alpha \alpha'} \delta( \varepsilon - \varepsilon' )\;,
\end{eqnarray}
where each Kronecker symbol $\delta_{xx'}$ implies that all quantum numbers in the sets $x$ and $x'$ are equal. The projection operators $\hQ$, $\hR$ and $\hP$ on the corresponding subspaces are given by the expressions
\begin{eqnarray}
\hQ &=& \sum_\beta  \ket{\beta} \bra{\beta} \;\label{Q_ketbra}\;,\\
\hR &=& \sum_f \int  d\omega \ket{f \omega} \bra{f \omega} \;, \label{R_ketbra}\\
\hP &=& \sum_\alpha \int  d\varepsilon   \ket{ \alpha \varepsilon} \bra{ \alpha \varepsilon }  \;, \label{D_ketbra}
\end{eqnarray}
and satisfy the orthogonality condition $\hQ\hR = \hR\hP =\hP\hQ = 0$. In the current description we neglect contributions from states lying outside the introduced subspaces and assume the completeness condition
\begin{equation}
\hQ + \hR + \hP = \mathds{1}\;,
\end{equation}
where $\mathds{1}$ is the identity operator. 

The Hamiltonian describing the dynamics of the system can be represented in the form $\hH = \hH_0 + \hV$. Here $\hV$ is the interaction term leading to transitions between states from different subspaces $Q$, $R$ and $P$, which consist of eigenstates of the unperturbed Hamiltonian $\hH_0$.  We note that $\hH_0$ incorporates interactions which don't mix the states from $Q$, $R$ and $P$,  such as static Coulomb interaction between the electrons and the nucleus.  In the following we use the Green's operators with the complex energy variable $z$ as the argument
\begin{eqnarray}
\hG_0(z)&=&(z-\hH_0)^{-1} \;,\label{G0_def}\\
\hG(z)&=&(z-\hH)^{-1} \;.\label{G_def}
\end{eqnarray}
They obey the Lippmann-Schwinger equation
\begin{equation}\label{LS_0}
\hG(z) = \hG_0(z) + \hG_0(z) \hV \hG(z)\;.
\end{equation}
We follow Ref.~\cite{HaanJacobs1989} and introduce an auxiliary operator $\hC = \hR + \hP$. By acting with $\hC$ on (\ref{LS_0}) from left and right, and inserting $\mathds{1}=\hC+\hQ$ between $\hV$ and $\hG(z)$, we obtain
\begin{eqnarray}
(z-\hH_0) \hQ \hG  \hC &=& \hQ \hV \hQ \hG  \hC +\hQ \hV \hC \hG  \hC \label{QGC_0} \;,\\
(z-\hH_0) \hC \hG  \hQ &=& \hC \hV \hQ \hG  \hQ + \hC \hV \hC \hG  \hQ \label{CGQ_0}  \;,\\
(z-\hH_0) \hQ \hG  \hQ &=& \hQ + \hQ \hV \hQ \hG  \hQ +\hQ \hV \hC \hG  \hQ\;, \label{QGQ_0} \\
(z-\hH_0) \hC \hG  \hC &=& \hC + \hC \hV \hQ \hG  \hC + \hC \hV \hC \hG  \hC\;. \label{CGC_0}
\end{eqnarray}
Introducing the operator
\begin{equation}\label{Pz_def}
\hPhi(z)= \hC [\hC ( z - \hH_0 - \hV ) \hC]^{-1} \hC
\end{equation}
we rewrite (\ref{QGC_0})---(\ref{CGQ_0}) in the form
\begin{eqnarray}
\hQ \hG(z) \hC = [\hQ \hG(z) \hQ] \hV [\hC \hPhi(z) \hC] \label{QGC_1}\;,\\
\hC \hG(z) \hQ=[ \hC \hPhi(z) \hC ] \hV [ \hQ \hG(z) \hQ ] \label{CGQ_1} \;,
\end{eqnarray}
which after substitution into~(\ref{QGQ_0})---(\ref{CGC_0}) give
\begin{eqnarray}
\hQ \hG(z) \hQ = \hQ [ \hQ (z - \hH_0 - \hLambda(z) ) \hQ ]^{-1} \hQ\;, \label{QGQ_1} \\
\hC \hG(z) \hC = \hC \hPhi(z) \hC [\mathds{1} + \hV \hQ \hG(z)\hC] \label{CGC_1}\;,
\end{eqnarray}
where
\begin{equation}\label{Lz_def}
\hLambda(z) = \hV + \hV \hC \hPhi(z) \hC \hV\;.
\end{equation}

The transition operator characterizing behaviour of the system subject to perturbation $\hV$, is given by
\begin{equation}\label{Tz_Lz}
\hT(z) = \hV + \hV \hG(z) \hV = \hLambda(z) + \hLambda(z) \hQ \hG(z) \hQ \hLambda(z) \;,
\end{equation}
which was obtained using~(\ref{QGC_1})---(\ref{CGC_1}). We assume that the excited nuclear state decays radiatively and include this radiative decay in the current description along with the LANEEC process itself.  Both the initial and the final states belong therefore to the substate $R$ and the whole process is described by the projection
\begin{equation}\label{Tz_Lz_proj}
\hR \hT \hR = \hR \hLambda \hR + [\hR \hLambda \hQ][\hQ \hG \hQ][\hQ \hLambda \hR]  \;,
\end{equation}
where the property $\hQ^2 = \hQ$ was used. The required projections of the $\hLambda(z)$ and $\hG(z)$ operators can be evaluated based on the projection
\begin{equation}\label{Phi_projections}
\hC \hPhi  \hC = \hP \hPhi  \hP + \hP \hPhi  \hR + \hR \hPhi  \hP + \hR \hPhi  \hR\;,
\end{equation}
which we obtain below.

In the following we assume $\hR \hV \hR = 0$, i.e. $\hV$ does not couple the subspace $R$ with itself. Then by rewriting~(\ref{Pz_def}) as
\begin{equation}
(z-\hH_0) \hC \hPhi \hC= \hC + \hC \hV \hC \hPhi \hC
\end{equation}
and acting with the operators $\hP$ and $\hR$ from left and right we find
\begin{eqnarray}
(z-\hH_0)\hP\hPhi \hP&=&\hP+\hP\hV\hP\hPhi \hP + \hP\hV\hR\hPhi \hP\;, \label{DPD_2}\\
(z-\hH_0)\hR\hPhi \hP&=&\hR\hV\hP\hPhi \hP \;, \label{RPD_2}\\
(z-\hH_0)\hP\hPhi \hR&=&\hP\hV\hP\hPhi\hR + \hP\hV\hR\Phi \hR\;, \label{DPR_2}\\
(z-\hH_0)\hR\hPhi \hR&=&\hR+\hR\hV\hP\hPhi \hR\;. \label{RPR_2}
\end{eqnarray}
By substituting $\hR \hPhi \hP$ from~(\ref{RPD_2}) into~(\ref{DPD_2}) we find
\begin{equation}\label{DPD_2.5}
\hP [z-\hH_0-\hV-\hV\hR\hG\hR\hV] [\hP\hPhi \hP] = \hP\;.
\end{equation}
Substitution of~(\ref{RPR_2}) into~(\ref{DPR_2}) and using (\ref{DPD_2.5}) gives
\begin{equation}\label{DPR_2.5}
\hP\hPhi \hR(z-\hH_0)=[\hP\hPhi \hP][\hP\hV\hR]\;.
\end{equation}
Expressing $\hP\hPhi \hP$ from~(\ref{DPD_2.5}), it is possible to find the other three projections required in (\ref{Phi_projections}) using~(\ref{RPD_2}),~(\ref{RPR_2}) and~(\ref{DPR_2.5}). We obtain in the following their matrix elements.

By taking matrix element from the operator equation~(\ref{DPD_2.5}) and insertion of the projection operators in the representation (\ref{Q_ketbra})---(\ref{D_ketbra}), we find
\begin{widetext}
\begin{eqnarray}\label{DPD_3}
(z-\varepsilon) \braket{ \alpha\varepsilon | \hPhi(z) | \alpha'\varepsilon'}
-\sum_{\alpha''} \int d\varepsilon'' \braket{ \alpha \varepsilon | \hV |  \alpha'' \varepsilon''} \braket{  \alpha'' \varepsilon'' | \hPhi(z) |  \alpha' \varepsilon' } \nonumber\\
-\sum_{\alpha''f} \iint d\varepsilon'' d\omega \frac{ \braket{  \alpha \varepsilon | \hV |f  \omega} \braket{ f \omega | \hV |  \alpha'' \varepsilon'' } \braket{  \alpha'' \varepsilon
'' | \hPhi(z) |  \alpha' \varepsilon'}} {z-\omega}
=\delta_{\alpha \alpha'} \delta(\varepsilon - \varepsilon')
\;.
\end{eqnarray}
\end{widetext}
We are interested in the limit $z=\lim_{\delta \to +0} ( \omega_p +i \delta )$ at some energy $\omega_p$. According to Sokhotski theorem \cite{Vladimirov_EqMathPh_1971},  the following operator equation holds
\begin{equation}
\lim_{\delta \to +0}\frac{1}{\omega_p+i\delta-\omega}= \left( \frac{1}{\omega_p-\omega}  \right)_\text{p.p.} -i\pi\delta ( \omega_p - \omega )\;.
\end{equation}
We omit the principal part (denoted by the subscript p.p.) adopting in this way the pole approximation. After substitution into~(\ref{DPD_3}), carrying out integration over $\omega$  explicitly and introducing auxiliary operators
\begin{eqnarray}
\hU &=&-i\pi \sum_f  \Bigl( \hV \ket{f \omega_p} \bra{ f \omega_p} \hV \Bigr) \;,\label{u2def} \\
\hW &=& \hV+\hU\;,\label{w2def}
\end{eqnarray}
we obtain
\begin{eqnarray}
&& \sum_{\alpha''} \int d \varepsilon''  \braket{ \alpha \varepsilon | z-\hH_0 - \hW | \alpha'' \varepsilon''}  \nonumber  \\
&&\times \braket{ \alpha'' \varepsilon'' | \hPhi(z) | \alpha' \varepsilon'}
= \delta_{\alpha \alpha'} \delta(\varepsilon-\varepsilon') 
\;. \label{DPD_3.7}
\end{eqnarray}
We solve this equation with respect to $\braket{ \alpha \varepsilon | \hPhi | \alpha' \varepsilon'}$ using the ansatz
\begin{equation}\label{Inv_ansatz}
\braket{ \alpha \varepsilon | \hPhi(z) | \alpha' \varepsilon'} = \frac{\delta_{\alpha \alpha'} \delta(\varepsilon-\varepsilon') }{z-\varepsilon' }+\frac{ \braket{ \alpha \varepsilon | \hphi(z) | \alpha' \varepsilon'} }{(z-\varepsilon)(z-\varepsilon')}\;,
\end{equation}
leading after substitution into~(\ref{DPD_3.7}) to the integral equation
\begin{eqnarray}
 &&\braket{ \alpha \varepsilon | \hphi(z) | \alpha' \varepsilon'} -  \braket{ \alpha \varepsilon | \hW | \alpha' \varepsilon'} \label{Inv_inteq_fi}\\
&& = \sum_{\alpha''} \int d\varepsilon'' \frac{ \braket{ \alpha \varepsilon | \hW | \alpha'' \varepsilon''} \braket{ \alpha'' \varepsilon'' | \hphi(z) | \alpha' \varepsilon'} }{z-\varepsilon''}\;. \nonumber
\end{eqnarray}
This in turn is solved with a power series expansion
\begin{equation}\label{Inv_expa} 
\braket{ \alpha \varepsilon | \hphi(z) | \alpha' \varepsilon'}
= \sum_{n=0}^\infty \braket{ \alpha \varepsilon | \hphi^{(n)}(z) | \alpha' \varepsilon'}\;,
\end{equation}
where $\hphi^{(n)}(z)$ denotes the term containing the $\hW$ operator $n$ times. We find $\hphi^{(0)}(z)=0$,
\begin{equation}\label{Inv_fi1}
\braket{ \alpha \varepsilon | \hphi^{(1)}(z) | \alpha' \varepsilon'}  = \braket{ \alpha \varepsilon | \hW | \alpha' \varepsilon'}\;,
\end{equation}
and for $n \geq 2$ after application of the pole approximation
\begin{eqnarray}
\braket{ \alpha \varepsilon | \hphi^{(n)}(z) | \alpha' \varepsilon'} = -i \pi \sum_{\alpha_1 \alpha_2} \braket{ \alpha \varepsilon | \hW | \alpha_1 \omega_p} 
\nonumber \\
\times \left[\tilde{X}^{n-2}\right]_{\alpha_1 \alpha_2} \braket{ \alpha_2 \omega_p | \hW | \alpha' \varepsilon'}
\end{eqnarray}
with the matrix $\tilde{X}$ defined as
\begin{equation}\label{Inv_Xtilde_def}
\tilde{X}_{\alpha_1 \alpha_2} = -i\pi \braket{ \alpha_1 \omega_p | \hW | \alpha_2 \omega_p}\;.
\end{equation}
Using the obtained powers $\hphi^{(n)}(z)$, we derive from (\ref{Inv_ansatz}) and (\ref{Inv_expa})
\begin{equation}\label{DPhiD}
\braket{ \alpha \varepsilon | \hPhi(z)  | \alpha' \varepsilon'} = -i\pi \delta(\varepsilon - \omega_p) \delta (\varepsilon' - \omega_p) \tilde{S}_{\alpha \alpha'}\;,
\end{equation}
where
\begin{equation}\label{Inv_Stilde_def}
\tilde{S} = \sum_{n=0}^\infty \tilde{X}^n \;.
\end{equation}

This result is now used to obtain the matrix elements of the other $\hPhi(z)$ operator projections needed in (\ref{Phi_projections}). The matrix element of the operator equation (\ref{RPD_2}) in the pole approximation leads to
\begin{eqnarray}
 \braket{f \omega | \hPhi(z) | \alpha \varepsilon} = (-i\pi)^2 \delta(\omega-\omega_p) \delta(\varepsilon-\omega_p) \nonumber \\ 
 \times\sum_{\alpha'} \braket{f \omega | \hV | \alpha' \omega_p} \tilde{S}_{\alpha' \alpha}\;.\label{RPhiD}
\end{eqnarray}
Analogously, from (\ref{DPR_2.5})
\begin{eqnarray}\label{DPhiR}
\braket{  \alpha \varepsilon | \hPhi(z) | f \omega } = (-i\pi)^2 \delta(\varepsilon-\omega_p) \delta(\omega-\omega_p) \nonumber\\ 
\times \sum_{\alpha'} \tilde{S}_{\alpha \alpha'}\braket{ \alpha' \omega_p | \hV |  f \omega } \;,\label{DPhiR}
\end{eqnarray}
and from (\ref{RPR_2})
\begin{widetext}
\begin{equation}\label{RPhiR}
\braket{ f \omega | \hPhi(z) | f' \omega' } = -i \pi  \delta(\omega-\omega_p) \delta(\omega'-\omega_p) \Bigl[
\delta_{f f'} 
+ (-i\pi)^2 \sum_{\alpha' \alpha''} \braket{f \omega_p | \hV |  \alpha' \omega_p } \tilde{S}_{\alpha' \alpha''}\braket{ \alpha'' \omega_p | 'hV |  f' \omega_p }
\Bigr]\;. \nonumber
\end{equation}
\end{widetext}
Using the definition of $\hLambda(z)$ given by (\ref{Lz_def}) with the projection operators rewritten in the representation (\ref{Q_ketbra})---(\ref{D_ketbra}), and substituting the obtained matrix elements of $\hPhi(z)$, we find
\begin{equation}\label{Lz_obt}
\hLambda(z) = \hW - i\pi \sum_{\alpha_1 \alpha_2} \hW \ket{\alpha_1 \omega_p} \tilde{S}_{\alpha_1 \alpha_2}\bra{\alpha_2 \omega_p} \hW\;.
\end{equation}
Matrix elements of the projection $\hQ \hG \hQ$ are then evaluated from (\ref{QGQ_1}) as
\begin{equation}\label{Eq_forG}
\sum_{\beta'}\braket{ \beta | z- \hH_0 -\hLambda(z) | \beta'} \braket{\beta' |\hG| \beta }= \delta_{\beta \beta'}\, .
\end{equation}
Since generally a finite number of bound states is involved in the process, the solution of this equation reduces to inversion of a finite-dimensional matrix.

We adopt in the following the so-called isolated resonance approximation by assuming $\braket{\beta |\hG| \beta' } = g_\beta \delta_{\beta \beta'}$. From (\ref{Eq_forG})
\begin{eqnarray}
g_\beta &=& \Bigl(z-E_\beta^{(0)} - \braket{\beta| \hW|\beta}  \label{Expr_gbeta}\\
			&+&  i\pi \sum_{\alpha_1 \alpha_2} \braket{\beta| \hW |\alpha_1 \omega_p} \tilde{S}_{\alpha_1 \alpha_2}\braket{\alpha_2 \omega_p| \hW|\beta} \Bigr)^{-1}\;,\nonumber
\end{eqnarray}
where $E_\beta^{(0)}$ is the eigenvalue of the unperturbed Hamiltonian $\hH_0$ corresponding to the state $\ket{\beta}$. As next step, we carry out the summation in the $\tilde{S}$ matrix definition (\ref{Inv_Stilde_def}) within the expression
\begin{equation}\label{Summation_what}
\sum_{\alpha_1} \braket{ \beta| \hW|\alpha_1 \omega_p} \tilde{S}_{\alpha_1 \alpha_2} =
\sum_{n=0}^\infty \left( \sum_{\alpha_1} \braket{ \beta| \hW|\alpha_1 \omega_p} \left[ \tilde{X}^n\right]_{\alpha_1 \alpha_2} \right)
\end{equation}
entering~(\ref{Expr_gbeta}). We make use of auxiliary operators $\hY_n$ defined recursively as
\begin{itemize}
\item $\hY_0 = \hW\;,$
\item $\hY_{n+1} = -i \pi \sum_\alpha \Bigl( \hY_n \ket{\alpha \omega_p} \bra{ \alpha\omega_p} \hW \Bigr)\;.$
\end{itemize}
From the definitions of $\hY_n$ and $\tilde{X}$ in (\ref{Inv_Xtilde_def}) follows the property
\begin{equation}
\sum_{\alpha_1} \braket{ \beta| \hY_n|\alpha_1 \omega_p} \tilde{X}_{\alpha_1 \alpha_2} = \braket{ \beta| \hY_{n+1}|\alpha_2 \omega_p}\;.
\end{equation}
The sum in~(\ref{Summation_what}) reduces then to
\begin{equation}
\sum_{\alpha_1} \braket{ \beta| \hW|\alpha_1 \omega_p} \tilde{S}_{\alpha_1 \alpha_2} = \Braket{ \beta|\sum_{n=0}^\infty \hY_n |\alpha_2 \omega_p}\;.
\end{equation}
Using this expression in (\ref{Expr_gbeta}), we find
\begin{equation}\label{gbeta_corrected}
g_\beta = \left(
z-E_\beta^{(0)} - \sum_{n=0}^\infty \braket{\beta| \hY_n|\beta}\;,
\right)^{-1}\;.
\end{equation}
where we used the property of the $\hY_n$ operators
\begin{eqnarray}
\braket{\beta| \hW|\beta} -  i\pi \sum_{\alpha}\Braket{ \beta|\sum_{n=0}^\infty \hY_n|\alpha \omega_p}
\nonumber
\\
\times
\braket{\alpha \omega_p| \hW|\beta} = \sum_{n=0}^\infty \braket{ \beta| \hY_n| \beta}
\end{eqnarray}
following from their definition.

The sum in~(\ref{gbeta_corrected}) contains energy corrections to the level $\ket{\beta}$ and its decay rates due to different mechanisms. Let us consider for example the contribution from the term $\braket{\beta| \hY_0 |\beta}=\braket{\beta| \hW|\beta}$, which according to the definitions (\ref{u2def}) and (\ref{w2def}) of the $\hU$ and $\hW$ operators can be written as
\begin{eqnarray}
\braket{\beta| \hY_0 |\beta} &=& \braket{\beta| \hV|\beta} - i\pi \sum_f \braket{\beta| \hV|f \omega_p}\braket{f \omega_p| \hV| \beta} \nonumber\\
&=& \braket{\beta| \hV|\beta} - \frac{i}{2}\left[ 2\pi \sum_f \left| \braket{\beta| \hV|f \omega_p} \right|^2 \right]\;.
\end{eqnarray}
The term $\braket{\beta| Y_0|\beta}$ accounts thus for the first-order energy shift caused by the perturbation $\hV$ and the radiative decay rate of the state $\ket{\beta}$ (in the brackets). The next term is
\begin{eqnarray}
\braket{\beta| \hY_1|\beta} = -i\pi \sum_\alpha \braket{\beta| \hW |\alpha \omega_p} \braket{\alpha \omega_p| \hW |\beta} \nonumber \\
= -i\pi \sum_\alpha \braket{\beta| \hV+\hU|\alpha \omega_p} \braket{\alpha \omega_p| \hV+ \hU|\beta}\, . \label{Y1_shift_decay}
\end{eqnarray}
The contribution in~(\ref{Y1_shift_decay}) containing only $\hV$ and not $\hU$ reads
\begin{equation}
-\frac{i}{2}\left[ 2\pi \sum_\alpha \left| \braket{\beta| \hV|\alpha \omega_p} \right|^2 \right]
\end{equation}
and thus represents the decay of the state $\ket{\beta}$ to states from the subspace $P$, i.e., Auger decay rate and IC. The other contributions from $\braket{\beta| \hY_n|\beta}$ at $n=1$ and higher $n$ are higher order corrections to the mentioned energy shift and decay rates. We write generically
\begin{equation}
g_\beta = \left(
z-E_\beta + \frac{i}{2} \Gamma_\beta
\right)^{-1}\;,
\end{equation}
where $E_\beta$ is the energy shifted by the interaction in all orders and $\Gamma_\beta$ is the total decay rate of the state $\ket{\beta}$. We note however that contributions due to the system states lying outside the considered subspaces are not included in $E_\beta$ and $\Gamma_\beta$.

We evaluate finally the matrix element of the transition operator projection (\ref{Tz_Lz_proj}). Using the results obtained above we obtain after algebraic simplifications
\begin{eqnarray}
&&\braket{f \omega | \hT | f' \omega'} = \braket{f \omega | \sum_{n=0}^\infty \hY_n | f' \omega'} \nonumber  \\
&&+ \sum_\beta
\frac{
\braket{f \omega | \sum_{n=0}^\infty  \hY_n| \beta} \braket{\beta | \sum_{n=0}^\infty \hY_n | f'\omega' }
}
{z-E_\beta + \frac{i}{2} \Gamma_\beta}\;.   \label{T_obt}
\end{eqnarray}
The first term describes processes that do not go through bound states $\ket{\beta}$ and are thus omitted in the current formalism giving finally
\begin{equation}\label{T_obt_LANEEC}
\braket{f \omega | \hT | f' \omega'} = \sum_\beta
\frac{
\braket{f \omega | \sum_{n=0}^\infty  \hY_n| \beta} \braket{\beta | \sum_{n=0}^\infty \hY_n | f'\omega' }
}
{z-E_\beta + \frac{i}{2} \Gamma_\beta}\;. 
\end{equation}

\section{Amplitude and rate of LANEEC process \label{Sec_1phot}}

The very general result obtained above in the pole and isolated resonance approximations,  describes the LANEEC process with subsequent emission of a photon in all orders.  We adopt here the lowest order approximation by retaining the minimal amount of terms still reflecting the process scenario:
\begin{eqnarray}
&&\braket{f \omega | \hT | f' \omega'} =  \sum_\beta
\frac{
\braket{f \omega |  \hY_0| \beta} \braket{\beta | \hY_1 | f'\omega' }
}
{z-E_\beta + \frac{i}{2} \Gamma_\beta}  \nonumber\\
&&=
-i\pi \sum_{\alpha\beta}
\frac{
\braket{f \omega |  \hW| \beta} \braket{\beta | \hW | \alpha \omega_p} \braket{\alpha \omega_p | \hW | f'\omega' }
}
{z-E_\beta + \frac{i}{2} \Gamma_\beta}\;.
\label{T_1phot}
\end{eqnarray}
From this expression, the amplitude $A_\mathrm{LANEEC}$ of the LANEEC process without inclusion of radiative decay of the final state can be written as
\begin{equation}\label{LANEEC_Amplitude_Feshbach}
\braket{\beta | \hT | f \omega}  = -i\pi \sum_{\alpha}
\braket{\beta | \hW | \alpha \omega_p} \braket{\alpha \omega_p | \hW | f\omega }\;.
\end{equation}

\newcommand{\hen}{\hat{H}_{en}}
\newcommand{\hnr}{\hat{H}_{nr}}
\newcommand{\her}{\hat{H}_{er}}
\newcommand{\henm}{\hat{H}_\text{magn}}
\newcommand{\hhyp}{\hat{H}_\mathrm{int}}

At this point, we take into account the actual form of the perturbation operator $\hV$ specific to the considered process. Generally $\hV$ can be represented as the sum of terms coupling the subspaces $P$, $Q$ and $R$ pairwise:
\begin{equation}\label{LANEEC_vv_new}
\hV=\hen + \hnr + \her\;.
\end{equation}
The chosen notations $\hen$, $\hnr$ and $\her$ represent the physical meaning of each operator: the Coulomb coupling of the electronic shell to the nucleus, the interaction of the nucleus with the radiation field, and interaction of the electrons with the radiation field, respectively. When substituting $\hV$ into Eq. (\ref{LANEEC_Amplitude_Feshbach}) via Eqs. (\ref{u2def})---(\ref{w2def}), however, only those terms from the general form (\ref{LANEEC_vv_new}) are kept, which reflect the LANEEC process scenario. We introduce also the magnetic interaction operator, describing interaction between the nucleus and the electrons via an intermediate virtual photon
\begin{equation}
\henm = -i \pi \hnr \sum_f \ket{f \omega_p} \bra{f \omega_p} \her \;.
\end{equation}
We obtain then the LANEEC amplitude in the form
\begin{equation}\label{LANEEC_Amplitude_Final}
\braket{\beta | \hT | f \omega}  = -i\pi \sum_{\alpha}
\braket{\beta | \hhyp | \alpha \omega_p} \braket{\alpha \omega_p| \her | f\omega }\;,
\end{equation}
where the operator $\hhyp = \hen + \henm$ represents the full interaction between the electrons and the nucleus. We note that the same operator $\hhyp$ describes the hyperfine structure of electronic levels.

We switch at this point to notations describing the system of interest more concretely and write the amplitude of the LANEEC process given by Eq. (\ref{LANEEC_Amplitude_Final}) in the form
\begin{eqnarray}\label{LANEEC_Amplitude_NewForm}
A_\mathrm{LANEEC}  &=& -i\pi 
\sum_{\kappa_c m_c}
\braket{ n_f \kappa_f m_f ; I_f M_f | \hhyp | \kappa_c m_c \varepsilon_c;I_i M_i} \nonumber \\
&\times& \braket{\kappa_c m_c \varepsilon_c | \her | n_i \kappa_i m_i}\;.
\end{eqnarray}
Here $I, M$ describe the nuclear total spin and its projection quantum numbers; $\kappa, m$ are the Dirac angular momentum quantum number (incorporating both the total angular momentum $j$ and the orbital angular momentum $l$) and the total angular momentum projection $m$ for the involved electron; $n$ denotes the principal quantum number for electronic bound states and $\varepsilon$ is the energy for continuum electronic states. The indices $i$ ($f$) correspond to the initial (final) nuclear or bound electronic states, whereas the index $c$ denotes quantities related to continuum electronic states.

We assume here the incident photons to be electric dipole photons polarized in $z$ direction. The operator $\her$ in Eqs. (\ref{LANEEC_Amplitude_NewForm}) reads then $\her = z E$ with the electric field $E$, which we treat classically.  The interaction operator $\hhyp$ in (\ref{LANEEC_Amplitude_NewForm}) is given by the scalar product
\begin{equation}
\hhyp = \sum_{kq} (-1)^{q} \hat{M}_{k,-q}  \hat{T}_{kq}\;,
\end{equation}
where $\hat{M}_{kq}$ and $ \hat{T}_{kq}$ are the spherical components of the nuclear multipole moment of rank $k$ and the respective electronic coupling operator (see e.g.~\cite{Johnson_book_2007}).  The LANEEC amplitude is then written as
\begin{eqnarray}\label{LANEEC_Amplitude_WithOperators}
&A_\mathrm{LANEEC} & = -i\pi E \sum_{kq} (-1)^{q} \braket{ I_f M_f | \hat{M}_{k,-q} | I_i M_i}
 \\
&\times&
 \sum_{\kappa_c m_c}
\braket{ n_f \kappa_f m_f | \hat{T}_{kq}| \kappa_c m_c \varepsilon_c } 
 \braket{\kappa_c m_c \varepsilon_c | z | n_i \kappa_i m_i}\;.\nonumber
\end{eqnarray}
Here we assume that the transition is excited by a broad band laser radiation with distribution $f$ such that its peak is tuned to the transition resonance.   The time-averaged rate of the LANEEC excitation is then obtained based on the amplitude as
\begin{equation}\label{LANEEC_Rate_Gen}
R_\mathrm{LANEEC} = \frac{2\pi f_\mathrm{max} (\tau_p \nu)}{2I_i+1}  \sum_{M_iM_fm_im_f} \left| A_\mathrm{LANEEC}  \right|^2 \;,
\end{equation}
where $f_\mathrm{max}$ is the maximal value of $f$, $\tau_p$ and $\nu$ are the pulse duration and repetition rate, respectively.  We sum here over the magnetic quantum numbers of the initial and final electronic orbitals,  since the former is assumed to be completely filled and the latter completely vacant prior to the LANEEC event. We also sum over the final and average over the initial magnetic substates.

The obtained expression can be applied to LANEEC with an additional hole (see the left and the right graphs in Fig. \ref{2photLANEEC_versions}) with the following corrections.  First, since the vacancies created by the first photon close very fast due to strong Auger decay channel,  some steady fraction $\alpha_h < 1$ of atoms in the sample possessing holes is needed for LANEEC.  Second,  the final electronic orbital is not completely vacant but possesses only one hole closed in the LANEEC event.  The time-averaged rate for the compound LANEEC process with an additional hole can be thus obtained based on $R_\mathrm{LANEEC}$ from Eq. (\ref{LANEEC_Rate_Gen}) as
\begin{equation}
R^\mathrm{+hole}_\mathrm{LANEEC} =  \frac{\alpha_h}{2j_f + 1} R_\mathrm{LANEEC}\;.
\end{equation}


\section{ Two-photon LANEEC \label{Sec_2phot}}

\begin{figure}[ht!]
\centering
\begin{tikzpicture}
\begin{feynman}
	\vertex (b) at (3, -1);
	\vertex (a) at (-1, -1);
	\vertex (c) at (4.5, -1);

	\vertex (d) at (-1, -2);
	\vertex [left=of d] (e);
	
	\vertex (f) at (-2.5, -4);
	\vertex [left=of f] (g);
	\vertex (h) at (3, -4);

	\diagram* {
		(a) -- [double, edge label'=$\ket{I_iM_i}$] (b) -- [double, edge label'=$\ket{I_fM_f}$] (c), 
		(h) --[fermion, edge label=$\ket{ n_f \kappa_f, -m_f}$] (f) -- [fermion, edge label=$\ket{n_i\kappa_i m_i }$] (d) -- [fermion, edge label=$\ket{\varepsilon_c \kappa_c m_c}$] (h),  
		(e) -- [photon, edge label=$\omega$] (d), 
		(g) -- [photon, edge label=$\omega_0$] (f), 
		(h) -- [photon,] (b), 		
	};
\end{feynman}
\end{tikzpicture}
\caption{Feynman-Goldstone diagram for the two-photon LANEEC process. The wavy lines show the external photons  with frequencies $\omega_0$ and $\omega$ and a virtual photon of the nuclear interaction with the atomic shell. The double line corresponds to excitation of the nucleus,  while the single straight lines correspond to the involved vacancy and electron.  The states are denoted by quantum numbers as introduced in the text.  Note the notation of the hole state via the quantum numbers of the missing electron. \label{diag_LANEEC}}
\end{figure}
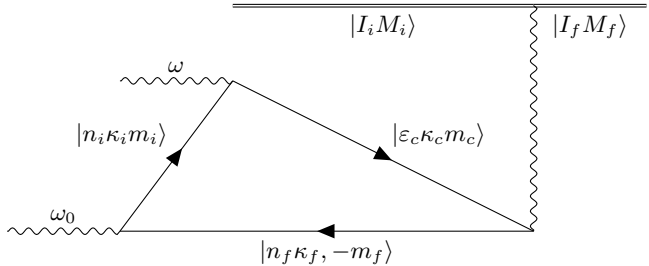

In the following we consider the two-photon LANEEC scenario introduced above (see the middle and the right graphs in Fig. \ref{2photLANEEC_versions}).  The amplitude of this process can be obtained based on its Feynman-Goldstone diagram shown in Fig. \ref{diag_LANEEC},  where the single and double solid lines represent electronic and nuclear states,  respectively,  and the wavy lines are the photon lines. 
A photon with the frequency $\omega_0$ creates a vacancy  in a deep inner shell.  The absorption of the photon with the frequency $\omega$ with subsequent electron-nucleus interaction is the LANEEC process described by the amplitude in Eq. (\ref{LANEEC_Amplitude_WithOperators}).  The amplitudes of the single photon excitation and the LANEEC process are coupled to a resulting amplitude via the equation
\begin{equation}\label{Ampl_2phot}
A^\mathrm{2phot}_\mathrm{LANEEC}  = \sum_{m_im_f} \frac{ A_\mathrm{LANEEC} \braket{ n_i\kappa_i m_i | Ez' |  n_f \kappa_f m_f}  }{  \omega_0 + E_f - E_i + \frac{i}{2}\Gamma_h } \;,
\end{equation}
where we introduce the energies of the corresponding electronic states $E_f$ and $E_i$, and assume for the additional photons linear polarization along the $Oz'$ axis, generally speaking different from the polarization axis $Oz$ in Eq.~(\ref{LANEEC_Amplitude_WithOperators}).  The energy width in the amplitude denominator reduces to the hole width $\Gamma_h$ only, since $\Gamma_h$ for deep inner electronic shells due to strong Auger decay channel exceeds significantly the width of the other (electronic and nuclear) states involved in the process.  Note that all electronic and vacancy states are intermediate  in this approach,  and imply thus summation over their magnetic quantum numbers in the amplitude.

The time-averaged rate is obtained based on the amplitude as
\begin{eqnarray}\label{LANEEC_2phot_Rate}
R^\mathrm{2phot}_\mathrm{LANEEC} &=& \int d \omega f_0(E_n -\omega) f(\omega) \\
&\times& \frac{2\pi(\tau_p \nu)}{2I_i+1}  \sum_{M_iM_f} \left| A^\mathrm{2phot}_\mathrm{LANEEC}  \right|^2 \;.
\end{eqnarray}
Here the integration is carried out over the distributions of the laser beams $f_0$ and $f$ assuming that the photon frequencies $\omega_0$ and $\omega$ match in total the nuclear transition energy $E_n$ and $\omega_0$ is tuned to the electronic transition between the bound states.  As before, we sum over the final and average over the initial nuclear magnetic substates, whereas the sum over the electronic magnetic substates enters directly the amplitude (\ref{Ampl_2phot}).

\section{ Numerical results \label{Sec_NR}}

In the following we show numerical examples for each case described above.  As an x-ray laser system we take the x-ray free-electron laser SACLA in Harima, Japan,  and assume the following radiation parameters \cite{SACLA_overview}.
\begin{table}[h!]
\centering
\renewcommand{\arraystretch}{1.3}
\begin{tabular}{|c|c|}\hline
Photon energy & $4-20$ keV \\\hline
Repetition rate & $\nu=30$~Hz \\\hline
Pulse duration & $\tau_p=10$~fs \\\hline
Spectral shape & Lorentzian \\\hline
Relative FWHM & $ \Delta \omega / \omega=0.5\%$ \\\hline
Intensity & $I=10^{18}\;\frac{\text{W}}{\text{cm}^2}$ \\\hline
Focal diameter & $d=1\;\mu\text{m}$ \\\hline
\end{tabular}
\caption{x-ray pulse parameters at SACLA assumed based on Ref.~\cite{SACLA_overview}. }
\label{EBFe_SACLA_params}
\end{table}
Required electronic matrix elements are evaluated based on the wave-functions obtained using the GRASP2K \cite{grasp} and RATIP \cite{ratip} packages for the bound and continuum states, respectively.  The bound transition energies are obtained with GRASP2K, if not otherwise stated.  Since no high precision is required, we restricted our GRASP calculations to the Multiconfiguration Dirac-Hartree-Fock model without additional electronic correlations.  The nuclear parameters were taken from the database \cite{NuclDB}. 

\subsection{ One-photon LANEEC }

As an example of ``pure'' one photon LANEEC process described in Section~\ref{Sec_1phot}, we consider LANEEC excitation in highly charged ions $\isotope[201][80]{Hg}^{44+}$ and $\isotope[205][82]{Pb}^{52+}$ with nuclear transitions of the $M1$ and $E2$ type,  respectively.  The energies $E_n$ of the transitions are 1.565 and 2.329 keV, respectively.  The strength of the coupling of the nuclear and electronic transitions is characterized by the internal conversion coefficient (ICC).  We choose therefore optimal charge states and electronic orbitals based on ICC obtained for all electronic shells from \cite{Kibedi_ICCoeff_2008}.  The information provided in this database for neutral atoms suffices for observation of the relative ICC behaviour in dependence of the involved electronic orbital also for higher charge states. 

In Table~\ref{LANEEC_1phot_beta} we show the ratio $\beta_\mathrm{LANEEC} = R_\mathrm{LANEEC} / R_\mathrm{rad}$ of the LANEEC rate to the rate of direct radiative excitation of the nucleus,  calculated based on Eq. (\ref{LANEEC_Rate_Gen}) for selected initial (Init.) and final (Fin.) orbitals for the LANEEC process.
\begin{table}[h!]
\centering
\renewcommand{\arraystretch}{1.3}
\begin{tabular}{|c|c|c|c|c|c|}\hline
\multirow{2}{*}{Ion} & \multirow{2}{*}{El. Conf. } & \multicolumn{2}{c|}{LANEEC} & \multirow{2}{*}{$\omega$, keV} &  \multirow{2}{*}{$\beta_\mathrm{LANEEC}$} \\ \cline{3-4}
&& Init. & Fin. && \\ \hline
\multirow{2}{*}{$\isotope[201][80]{Hg}^{44+}$} & \multirow{2}{*}{$[\isotope{Ar}]3d^{10}4s^24p^6$} & $3p_{3/2}$ & $5s$ & 5.02 & $1.5 \cdot 10^{-3}$ \\ \cline{3-6}
&& $3p_{1/2}$ & $5s$ &5.47 & $4.0 \cdot 10^{-4}$ \\ \hline
\multirow{2}{*}{$\isotope[205][82]{Pb}^{52+}$}& \multirow{2}{*}{$[\isotope{Ar}]3d^{10}4s^2$} & $4s$ & $5p_{1/2}$ & 3.70 & 12 \\ \cline{3-6}
&& $4s$ & $5p_{3/2}$ & 3.70 & 24 \\ \hline
\end{tabular}
\caption{The ratio $\beta_\mathrm{LANEEC} = R_\mathrm{LANEEC} / R_\mathrm{rad}$ for the one-photon LANEEC rate to the rate of direct radiative nuclear excitation.  The electronic configurations are shown with respect to the argon core configuration $1s^22s^22p^63s^23p^6$.  See the text for further explanations.}
\label{LANEEC_1phot_beta}
\end{table}
In the case with $\isotope[205][82]{Pb}^{52+}$ we obtained a few orders larger enhancement $\beta_\mathrm{LANEEC}$ than for $\isotope[201][80]{Hg}^{44+}$.  This is explained by significantly lower direct radiative excitation rate of the former transition due to its $E2$ type and low nuclear transition energy.  We observe that only very moderate enhancement due to involvement of the electronic shell is achieved in the ``pure'' one-photon LANEEC process.  As already mentioned, further nuisance is that this LANEEC version requires photons of higher energies than the nuclear transition energy.  We discuss in the following improved LANEEC schemes which may be of interest for future applications, since they allow for both more pronounced advantage with respect to the direct excitation,  and extension of addressed nuclear transitions to higher energies.

\subsection{LANEEC with additional hole}

As a potentially useful application, we consider here excitation of the 29.2~keV nuclear state in $\isotope[229][90]{Th}$.  The $\isotope[229][90]{Th}$ isomer is of interest due to its very low lying isomeric state at approx.  0.01~keV \cite{Wense_Nature_2016, Seiferle_Nature_2019, Sikorsky_PRL_2020, Yamaguchi_PRL_2019},  which can be used e.g. for implementation of the first nuclear clock at an unprecedented accuracy \cite{Peik_Clock_2003,  Peik_Review_2020}.  A possible (indirect) isomer excitation mechanism demonstrated in Refs. \cite{Yamaguchi_PRL_2019, Masuda_Nature_2019} employs excitation of the 29.2~keV nuclear state with high-brilliance synchrotron radiation. The latter decays then predominantly to the isomeric state.

We study here the possibility to excite the 29.2~keV level using the LANEEC process in neutral $\isotope[229][90]{Th}$ atoms.  Since this energy is not achievable at the SACLA facility, we consider a modified version of LANEEC,  in which the final electronic state is not in an outer shell,  but in a vacancy created in a deep-lying closed shell by another x-ray photon (see the left and the right graphs in Fig. \ref{2photLANEEC_versions}).  In this way,  the considered process involves two photons, but differs from the two-photon LANEEC excitation described in Section~\ref{Sec_2phot}. Here the first incoming photon only expels an electron from the deep-lying shell, which leaves the atom and does not further participate in the process. The sum of the two photon energies in this case does not need to be equal to $E_n$.

As a concrete implementation, we consider the scheme from Ref.~\cite{BilousThesis2018} with two SACLA beams at energies 20.8 and 8.6 keV irradiating a $\isotope[229][90]{Th}$ sample. The first beam creates vacancies in the $2s$ shells in the sample atoms, whereas the second one induces the one-photon LANEEC process as considered in Section~\ref{Sec_1phot}. At the latter stage a $6p$ electron is promoted to a continuum state which decays then into the $2s$ vacation with simultaneous excitation of the nucleus. The energy of the second photon is chosen such that the needed energy of 29.2~keV is transferred to the nucleus.

Vacancies in the $2s$ shell close very fast due to strong electronic Auger decay resulting in the width of the hole state of $\Gamma_h \approx 14.3 \;\mathrm{eV}$ \cite{ThHoleLifetime_PRA} corresponding to the lifetime $\tau_h \approx 50 \;\mathrm{as}$. Using the photoionization cross section $\sigma_h \approx 5.0 \;\mathrm{kb}$ calculated theoretically in Ref.~\cite{Scotfield_Photoion_Crossect_Tables}, we find that the steady time-averaged fraction $\alpha_h \approx 7.5 \cdot 10^{-5}$ of atoms have a vacancy in the $2s$ shell.  The excitation rate per atom in this compound process can be obtained as $R^\mathrm{+hole}_\mathrm{LANEEC} = \alpha_h R_\mathrm{LANEEC}$, where the latter rate is given by Eq.~(\ref{LANEEC_Rate_Gen}). 

For the reduced transition probabilities $B_\downarrow$ for the $M1+E2$ transition from the 29.2 keV level to the ground state we use the values $B_\downarrow(M1) = 0.003 \;\mathrm{W.u.}$ and $B_\downarrow(E2) = 27.11 \;\mathrm{W.u.}$ based on nuclear structure calculations performed in Refs. \cite{Minkov_Palffy_PRC_2021, Kirschbaum2022_PRC}.  The calculated nuclear excitation rate is $R^\mathrm{+hole}_\mathrm{LANEEC} = 3 \cdot 10^{-16}\;\mathrm{s}^{-1}$.  For a $\isotope[229][90]{Th}$ sample of thickness 1 $\mu$m the number atoms exposed to the laser radiation is $2.4 \cdot 10^{10}$ leading in total to approx.  4 excitation events per week.  Although this number is very small and the scheme is not practically applicable at the moment, we would like to point out large enhancement with respect to direct two-photon excitation of the nucleus.  Our calculation shows that direct excitation using two 14.6 keV SACLA beams with parameters listed in Table~\ref{EBFe_SACLA_params} has the rate $R^\mathrm{2phot}_\mathrm{rad} = 1 \cdot 10^{-26} \;\mathrm{s}^{-1} $ per atom.  This yields the enhancement factor $\beta^\mathrm{+hole}_\mathrm{LANEEC} = 2\cdot 10^{10}$. 

In the considered approach, the energy of the photon ionizing a deeply lying shell is not strictly fixed and has only to exceed the ionization threshold.  This property allows excitation schemes with only one laser beam for both creation of a vacancy and inducing the LANEEC process.  Another advantage is that the rate becomes 4 times larger since the photon exchange term contributes in the amplitude in the same way as the direct term.  As an example, we consider here excitation of the same 29.2 keV nuclear level with creation of a hole in the $2p_{3/2}$ shell,  the $3s$ orbital as the starting point for the electronic path in LANEEC which ends in the created hole. Both steps are induced by a single SACLA beam at energy 16.5 keV. The lifetime of the hole is 80~as based on the width provided in Ref.~\cite{ThHoleLifetime_PRA}, the photoionization cross section is 25~kb~\cite{Scotfield_Photoion_Crossect_Tables},  leading to the steady hole fraction of $7.6 \cdot 10^{-4}$. The calculated excitation rate per atom is in this case $R^\mathrm{+hole}_\mathrm{LANEEC} = 2 \cdot 10^{-14}\;\mathrm{s}^{-1}$ and corresponds for a sample of 1 $\mu$m thickness to approx.  38 excitation events per day making the scheme though challenging today but interesting for future applications.  The enhancement with respect to direct two photon excitation is $\beta^\mathrm{+hole}_\mathrm{LANEEC} = 2\cdot 10^{12}$.  The rate $R^\mathrm{+hole}_\mathrm{LANEEC}$ and the enhancement $\beta^\mathrm{+hole}_\mathrm{LANEEC}$ are considerably larger than in the previous example mainly due to presence of a strong $E2$ channel. 

\subsection{ Two-photon LANEEC }

In this Section, we obtain the rate of the LANEEC process with two photons as described in Section~\ref{Sec_2phot}.  As an example,  we consider the 14.4~keV M\"ossbauer transition in the $\isotope[57]{Fe}$ nucleus.  Direct one-photon XFEL excitation of this transition has been recently achieved at the SACLA facility \cite{Chumakov2018}.  Here we consider a scenario involving the electronic shell and two photons of energies 7.1 and 7.3~keV at SACLA.  The 7.1~keV photon excites a $1s$ electron to the vacant $4p$ orbital.  The 7.3~keV photon promotes this electron to a continuum state,  which decays back into the $1s$ vacancy with transferring the energy $7.1+7.3=14.4$~keV to the nucleus.

The excitation rate is obtained using Eqs. (\ref{Ampl_2phot})---(\ref{LANEEC_2phot_Rate}) assuming plane polarization for both photons.  As before,  we consider $z$-direction for the electric vector in the beam inducing the LANEEC part of the process, whereas for the beam exciting the electron from the $1s$ to the $4p$ shell $x$-polarization is assumed, i.e. $z'=x$ in Eqs. (\ref{Ampl_2phot}).  The width of the $1s$ vacancy in Eq. (\ref{Ampl_2phot}) is determined predominantly by Auger decay and has the value $\Gamma_h = 1.2$ eV \cite{Campbell_Hole_Param_Tables}. Using the parameters of the XFEL beams presented in Table~\ref{EBFe_SACLA_params}, we obtain the rate per atom $R^\mathrm{2phot}_\mathrm{LANEEC} = 9 \cdot 10^{-21}\;\mathrm{s}^{-1}$.  The obtained direct two-photon excitation rate with plane polarization in the same direction in both beams is in this case $R^\mathrm{2phot}_\mathrm{rad} = 7 \cdot 10^{-25} \;\mathrm{s}^{-1} $ per atom leading to the enhancement factor $\beta^\mathrm{+hole}_\mathrm{LANEEC} = 1\cdot 10^{4}$.

We observe an interesting cancellation effect if both beams are polarized in $z$-direction,  i.e. for $z'=z$.  In this case the excitation rate turns out to be identically zero.  This peculiarity can be explained by applying the Wigner-Eckart theorem to the electronic matrix elements constituting the amplitude in Eq. (\ref{Ampl_2phot}). The summation over the intermediate magnetic quantum numbers reduces then to a summation with corresponding Clebsch-Gordan coefficients.  For the considered electronic states and photon polarizations this sum turns out to be zero.  This effect could be used in this case for switching the nuclear excitation on and off by changing the photon polarization.  Note however,  that the same effect takes place for a purely electronic process, in which the continuum electronic state decays into the $1s$ vacancy with emission of a photon.  Due to this reason further checks are necessary for unambiguous detection of the nuclear excitation. This is however an ubiquitous aspect in all NEEC-related considered processes.

\section{ Conclusions\label{Sec_Concl} }

In this work we develop a theoretical description of the LANEEC process based on the Feshbach projection operator formalism.  Numerical examples for experimental scenarios at the SACLA facility are provided.  The decay channels appear to all orders in a natural and unified manner in the developed formalism.  The ``pure'' LANEEC version involving one photon requires usage of x-ray beam energies higher than the nuclear transition energy.  The achieved enhancement with respect to direct excitation is at the same time very moderate (see Table \ref{LANEEC_1phot_beta}).  Due to these reasons we consider two improved LANEEC versions with an additional x-ray photon, referred to as ``LANEEC with additional hole'' and ``two-photon LANEEC'' (see Fig. \ref{2photLANEEC_versions} and explanations in the text).  Based on these schemes we describe experimental scenarios for excitation of the 29.2~keV nuclear state in $\isotope[229]{Th}$ and the 14.4 keV M\"ossbauer transition in $\isotope[57]{Fe}$ which are of interest for further applications.  Our calculations show low excitation rates but strong enhancement with respect to the direct two photon excitation.  These results are insightful and the developed formalism will be useful also for other excitation processes, despite the very challenging practical implementation of LANEEC.  First experimental efforts towards observation of the LANEEC process in $\isotope[57]{Fe}$ were undertaken at LCLS \cite{DavidReis_Private}.
 
We are grateful to A. ~P\'alffy for extensive discussions of the results and careful revision of the manuscript.  We thank D. ~Reis, A. ~Kaldun and J.~Haber for very useful discussions of the experimental aspects of LANEEC. 

\bibliography{refs}

\end{document}